\documentclass[12pt]{article}
\usepackage{graphicx}
\newcommand\pubdate{\today}
\newcommand\pubnumber{EFI 11-18}
\def\Title#1{\begin{center} {\Large #1 } \end{center}}
\def\Author#1{\begin{center}{ \sc #1} \end{center}}
\def\Address#1{\begin{center}{ \it #1} \end{center}}

\newcommand\pubblock{\rightline{\begin{tabular}{l} \pubnumber\\
         \pubdate  \end{tabular}}}
\newenvironment{Abstract}{\begin{center}{\bf Abstract}\end{center} \bigskip \begin{quotation}  }{\end{quotation}}
\newenvironment{Presented}{\begin{quotation} \begin{center} 
             PRESENTED AT\end{center}\bigskip 
      \begin{center}\begin{large}}{\end{large}\end{center} \end{quotation}}
\def\Acknowledgements{\bigskip  \bigskip \begin{center} \begin{large}
             \bf ACKNOWLEDGEMENTS \end{large}\end{center}}

\def\beq{\begin{equation}}
\def\eeq#1{\label{#1}\end{equation}}
\def\eeqn{\end{equation}}

\def\beqa{\begin{eqnarray}}
\def\eeqa#1{\label{#1}\end{eqnarray}}
\def\eeqan{\end{eqnarray}}

\let\bar=\overbar

\def\Dslash{\not{\hbox{\kern-4pt $D$}}}
\def\dslash{\not{\hbox{\kern-2pt $\del$}}}

\def\msb{{\bar{\ssstyle M \kern -1pt S}}}

\def \bea{\begin{eqnarray}}
\def \beq{\begin{equation}}
\def \brf{{\cal B}}
\def \eea{\end{eqnarray}}
\def \eeq{\end{equation}}
\begin{document}
\begin{titlepage}
\pubblock

\vfill


\Title{Charm at Threshold}
\vfill
\Author{Jonathan L. Rosner}  
\Address{Enrico Fermi Institute, University of Chicago, Chicago, IL 60637, USA}
\vfill


\begin{Abstract}

Results from the CLEO Collaboration, mainly dealing with the study of
charmed mesons produced at flavor threshold but also covering other areas of
CLEO's investigation, are reviewed.

\end{Abstract}

\vfill

\begin{Presented}
The Ninth International Conference on\\
Flavor Physics and CP Violation\\
(FPCP 2011)\\
Maale Hachamisha, Israel,  May 23--27, 2011
\end{Presented}
\vfill

\end{titlepage}
\def\thefootnote{\fnsymbol{footnote}}
\setcounter{footnote}{0}
%

\section{Introduction}

The CLEO Collaboration at the Cornell Electron Storage Ring (CESR) studied
$e^+ e^-$ collisions for nearly 30 years, taking its last data in March 2008.
During this time it accumulated a wealth of data on charm and bottom quarks,
some of which is still being analyzed.  This report contains a selection of
recent results.

In Sec.\ \ref{sec:det} we compare the properties of the CLEO III and
CLEO-c detectors, and describe data samples and analyses in Sec.\
\ref{sec:samples}.  An update of $D^0$ and $D^+$ branching fractions
(Sec.\ \ref{sec:bf}) is in its final stages of analysis.  Using the
correlated nature of $D^0$ mesons produced in $e^+ e^- \to \psi(3770)
\to D^0 \overline{D}^0$, one can study CP- and flavor-tagged Dalitz
plots for $D^0$ decays (Sec.\ \ref{sec:dp}), which can help in the
determination of the weak phase $\gamma/\phi_3$ in $B$ decays.

CLEO has recently performed an improved measurement of $\brf[\psi(2S) \to
\pi^0 h_c]$ (Sec.\ \ref{sec:hcbf}) and searched for the transition
$\Upsilon(3S) \to \pi^0 h_b$ search (sec.\ \ref{sec:hb}).  An unexpected result
was the large production cross section for $e^+ e^- \to \pi^+ \pi^- h_c$ at
$E_{\rm cm} = 4170$ MeV (Sec.\ \ref{sec:hcpr}).  Other recent CLEO results have
been obtained for charmed particle final states with leptons (Sec.\
\ref{sec:sl}), including a search for $D_s^+ \to \omega e^+ \nu_e$ (Sec.\
\ref{sec:omenu}).  Sec.\ \ref{sec:concl} concludes.

\section{CLEO III and CLEO-c detectors \label{sec:det}}

\begin{figure}
\includegraphics[width=4in,angle=270]{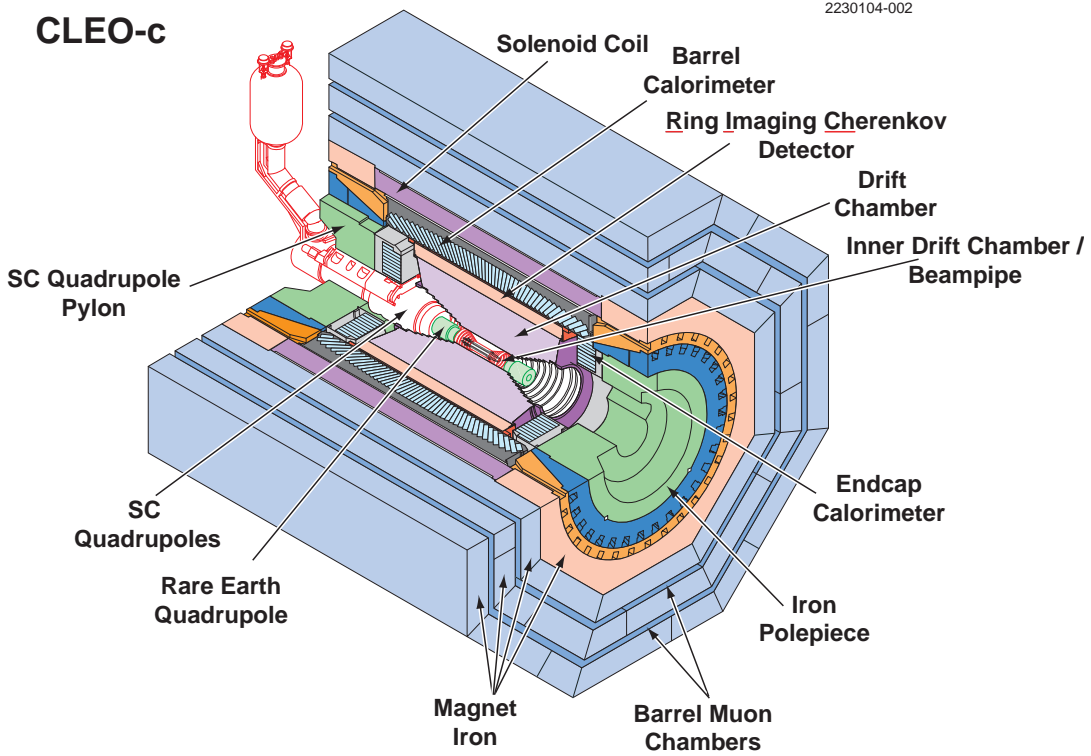}
\caption{The CLEO-c detector \label{fig:det}}
\end{figure}

Fig.\ \ref{fig:det} shows the latest incarnation of the CLEO detector,
known as CLEO-c.  It is the version used to study charmed meson production
near threshold.  It utilized an inner drift chamber, which replaced a
silicon vertex detector in the CLEO III version in order to reduce
multiple scattering.  The CLEO III detector was used for the study of
bottomonium and $B$ meson pair production.
Both versions of the detector had excellent neutral and charged particle
energy/momentum resolution:  $\Delta E/E = (5\%,2.2\%)$ at (0.1,1) GeV for
photons and $\Delta p/p = 0.6\%$ at 1 GeV for charged tracks.

\section{Data samples and analyses
\label{sec:samples}}

We will not report today on the pioneering CLEO results above bottom pair
production threshold, which include the first observation of the $B^0 = \bar b
d$ and $B^+ = \bar b u$ mesons.  Lower-energy samples discussed today are
summarized in Table \ref{tab:samples}.  Additional off-resonance samples were
taken for continuum studies.  The total number of CLEO publications as of
April 2011 was 517, with three more under review.  Approximately two dozen
other analyses were still in progress.  There have been 236 CLEO Ph.\ D.
degrees, with an additional 32 devoted to the physics of the CESR storage ring
and 14 more from the CUSB (Columbia University -- Stony Brook) group whose
detector operated on the other side of the ring from CLEO during the 1980s.

\begin{table}
\caption{CLEO data samples discussed in the present report.  $\psi(4170)$
denotes running at $E_{\rm cm} = 4170$ MeV; the accepted mass for the
$2^3D_1$ charmonium resonance is 4160 MeV/$c^2$ \cite{Nakamura:2010zzi}.
\label{tab:samples}}
\begin{center}
\begin{tabular}{c c c c} \\ \hline \hline
           & Initial & Integrated  & Decay \\
       & state & luminosity (pb$^{-1}$) & products \\ \hline
Charmonium  & $\psi(3686)$ & 53.8  & 27 M total \\
and charm   & $\psi(3770)$ & 818 & 5.3 M $D \bar D$ \\
            & $\psi(4170)$ & 586 & 0.6 M $D_s^+ D_s^{*-}$ + c.c. \\
Bottomonium & $\Upsilon(1S)$ & 1056 & $20.81$ M total \\
            & $\Upsilon(2S)$ & 1305 & $9.32$ M total \\
            & $\Upsilon(3S)$ & 1387 & $5.88$ M total \\ \hline \hline
\end{tabular}
\end{center}
\end{table}

\section{$D^0$, $D^+$ absolute branching fractions\label{sec:bf}}

\begin{figure}
\begin{center}
\includegraphics[height=5in]{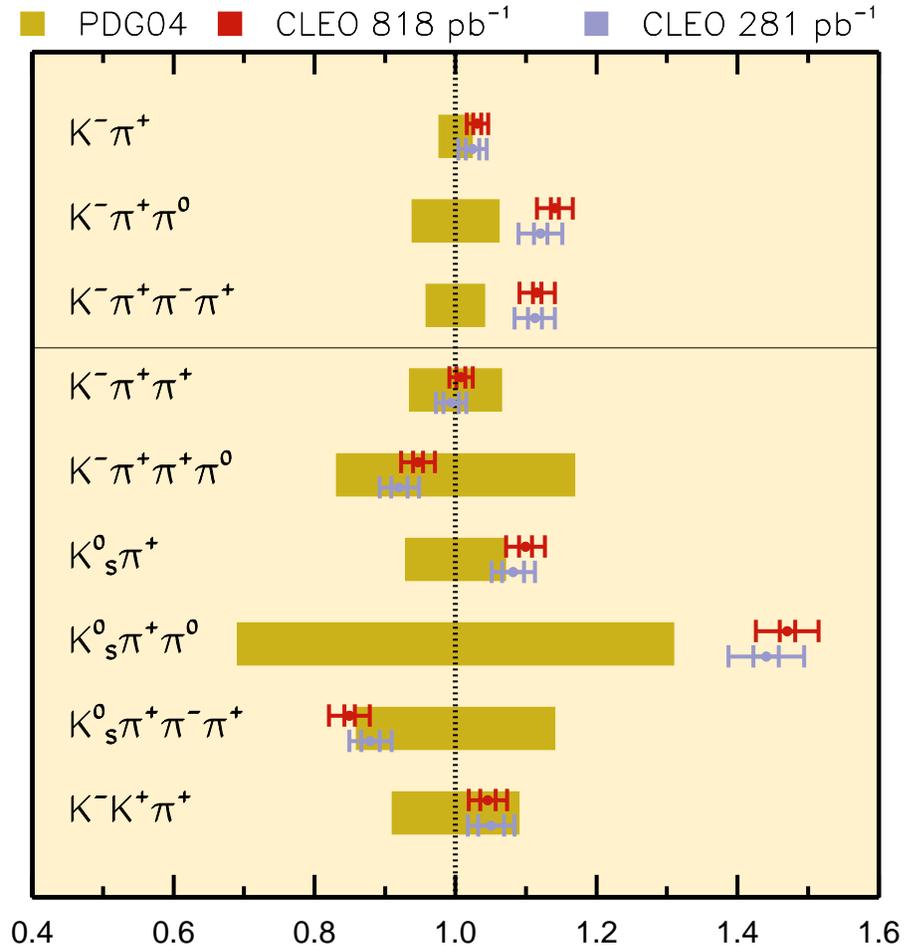}
\end{center}
\caption{Ratios of branching fractions measured by CLEO with respect to 2004
compilation by Particle Data Group \cite{Eidelman:2004wy}.  These results
are preliminary.
\label{fig:bfs}}
\end{figure}

Preliminary branching fractions for a number of $D^0$ and $D^+$ final states
based on the full sample of 818 pb$^{-1}$ \cite{xinshi} are compared in Fig.\
\ref{fig:bfs} with pre-CLEO-c Particle Data Group averages
\cite{Eidelman:2004wy} and with published CLEO results based on 281 pb$^{-1}$
\cite{Dobbs:2007zt}.  The analysis uses a combination of single-tag and
double-tag methods.  One outlying branching fraction is
${\cal B}(D^+ \to K_S \pi^+ \pi^0)$, found to be about 1.47 times the PDG 2004
value.

\section{Tagged Dalitz plots; weak phase $\gamma$\label{sec:dp}}

\begin{figure}
\begin{center}
\includegraphics[width=0.7\textwidth]{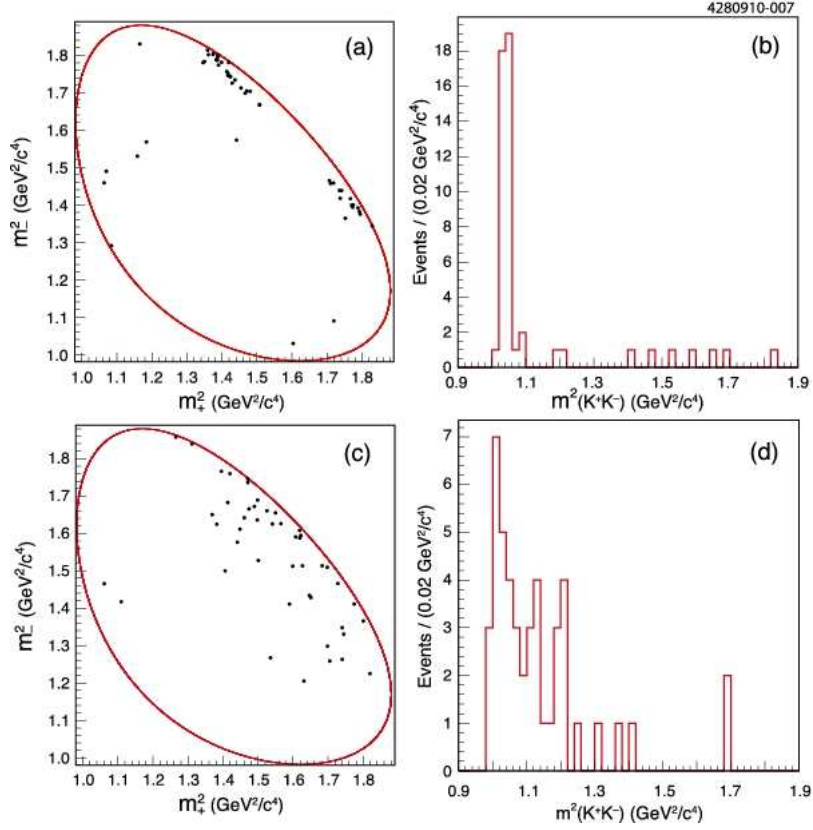}
\end{center}
\caption{Dalitz plots for $D \to K_S K^+ K^-$ and their $K^+ K^-$ projections.
Top:  CP-even tag ($\phi$ visible); bottom: CP-odd tag (no $\phi$).
\label{fig:KSKK}}
\end{figure}

The strong phase difference between $D^0$ and $\bar D^0$ decays to $K_{S,L} h^+
h^-$ ($h = \pi$ or $K$) is important to learn the Cabibbo-Kobayashi-Maskawa
(CKM) angle $\gamma/\phi_3$ in $B^- \to K^- D$ decays; the potential was seen
of reducing the error $\Delta \phi \simeq 9$--$10^\circ$ to a value of
3--4$^\circ$ using the current CLEO data \cite{Libby:2010nu} by exploting
the quantum coherence in $\psi(3770) \to D^0 \bar D^0$.  This goal has in
fact been achieved in a recent Belle analysis \cite{:2011vf}.

\begin{figure}
\begin{center}
\includegraphics[width=0.7\textwidth]{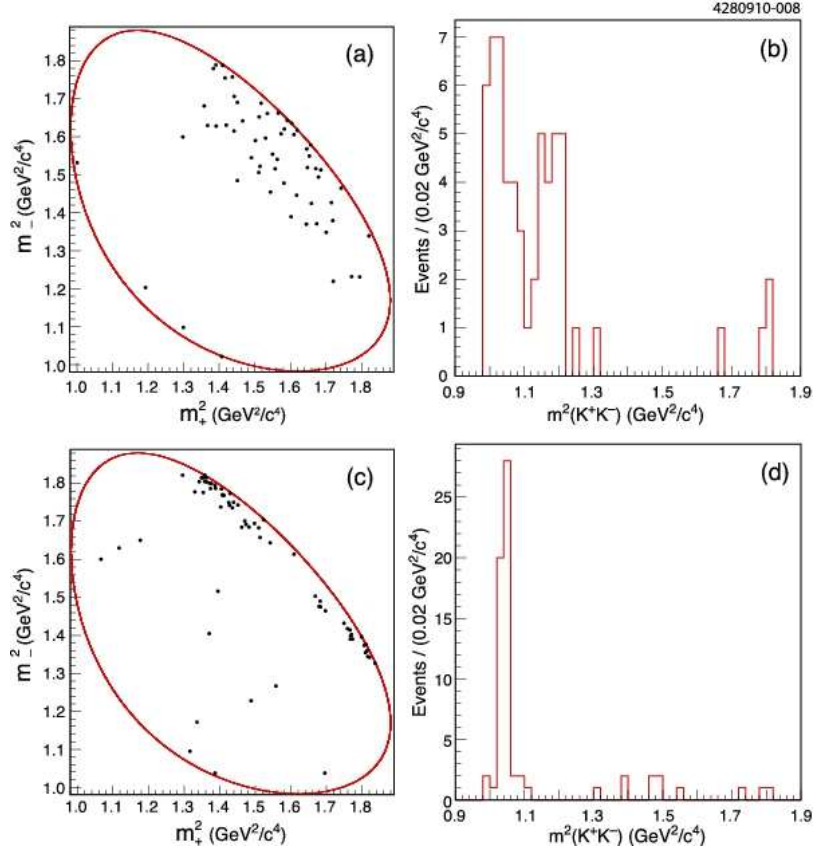}
\end{center}
\caption{Dalitz plots for $D \to K_L K^+ K^-$ and their $K^+ K^-$ projections.
Top:  CP-even tag (no $\phi$); bottom: CP-odd tag ($\phi$ visible).
\label{fig:KLKK}}
\end{figure}

As an illutration of this coherence we compare Dalitz plots for $D \to K_S K^+
K^-$ (Fig.\ \ref{fig:KSKK}) and $D \to K_L K^+ K^-$ (Fig.\ \ref{fig:KLKK}) in
which the companion $D$ (the ``tagging $D$'') in $\psi(3770)$ is CP-even (top
panels) or CP-odd (bottom panels).  The $\psi(3770)$ decay produces a pair of
$D$ mesons in opposite-CP states, so a CP-even tag will lead to a CP-odd
$K_{S,L} K^+ K^-$ final state, and vice versa.  This is borne out by the
presence of a $\phi$ in the $K^+ K^-$ spectrum in the CP-even-tagged $K_S
K^+ K^-$ and CP-odd-tagged $K_L K^+ K^-$ final states, but not in the other
two final states.

The use of $D^0$ Dalitz plots in the study of $B$ decays proceeds as follows
\cite{Naik:2011}.  The amplitudes ${\cal A}(B^- \to D^0 K^-) \sim V_{cb}
V^*_{us}$ and ${\cal A}(B^- \to \bar D^0 K^-) \sim V_{ub} V_{cs}$ can
interfere when they lead to the same final state $f(D)$.  Their ratio is
\beq
\frac{{\cal A}(B^- \to \bar D^0 K^-)}
                  {{\cal A}(B^- \to D^0 K^-)} = r_B e^{i(\delta_B - \gamma)}~,
~~r_B \equiv \left| \frac{{\cal A}(B^- \to \bar D^0 K^-)}
                  {{\cal A}(B^- \to D^0 K^-)} \right| \simeq 0.1~.
\eeq
where $\delta_B$ is a strong phase difference.
Atwood, Dunietz, and Soni \cite{Atwood:1996ci} proposed the use of $f(D) = K^+
\pi^-$, while Atwood and Soni \cite{Atwood:2003mj} suggested using multi-body
$D$ final states.  In the latter case one has to determine a {\bf coherence
factor} $R_F$ defined for the final state $F$ ($0 \le R_F \le 1$):
\beq
R_F e^{i \delta_D^F}=\frac{\int {\cal A}(s) \bar {\cal A}(s)
e^{i \delta(s)} ds}{\sqrt{\int |{\cal A}(s)|^2 ds \int |\bar {\cal
A}(s)|^2 ds}}~.
\eeq
Table \ref{tab:coh} shows some coherence factors determined by CLEO
\cite{Naik:2011}.
One can reduce model-dependence by performing fits to Dalitz plots with
bins of equal $\Delta \delta_D^F$ \cite{Giri:2003ty,Bondar:2005ki}.

\begin{table}
\caption{Coherence factors $R_F$ for final states $F$ in $D^0$ decays.
\label{tab:coh}}
\begin{center}
\begin{tabular}{c c c c} \hline \hline
\null \quad $F$ & $K \pi \pi^0$ & $K 3 \pi$ & $K_S K \pi$ \\ \hline
$R_F$ & $0.84 \pm 0.07$ & $0.33^{+0.26}_{-0.23}$ & $0.73 \pm 0.09$ \\
$\delta_D^F$ & $(227^{+14}_{-17})^\circ$ & $(114^{+26}_{-23})^\circ$
 & $(8.2 \pm 15.2)^\circ$ \\ \hline \hline
\end{tabular}
\end{center}
\end{table}

\section{${\cal B}[\psi(2S) \to \pi^0 h_c]$ \label{sec:hcbf}}

\begin{figure}
\begin{center}
\includegraphics[width=0.98\textwidth]{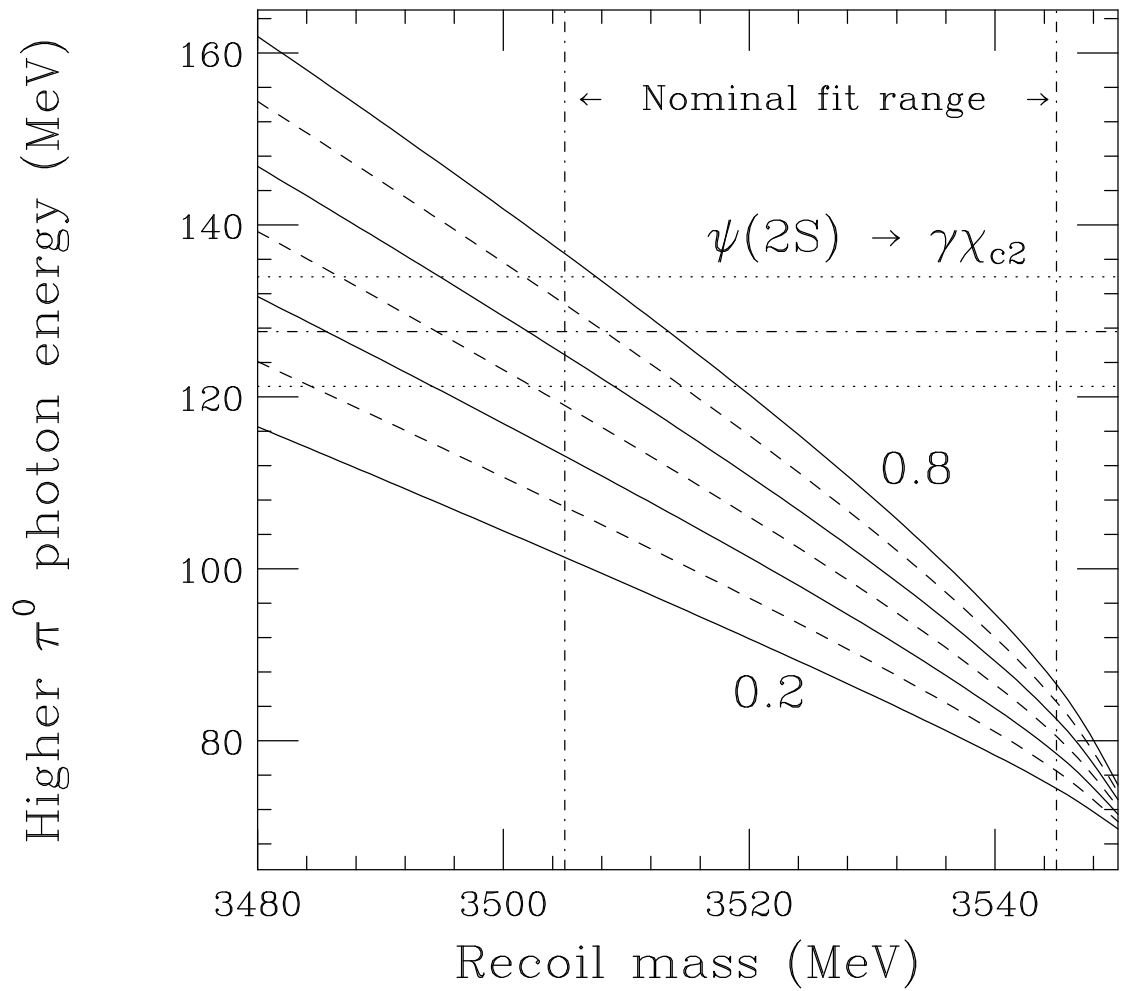}
\end{center}
\caption{Higher photon energy in $\pi^0 \to \gamma \gamma$ vs.\ recoil mass
$M(X)$ in $\psi(2S) \to \pi^0 X$, for various values of $|\cos \alpha|$, where
$\alpha$ is the $\pi^0$ decay angle (see text).
\label{fig:egam}}
\end{figure}

\begin{figure}
\begin{center}
\includegraphics[width=0.98\textwidth]{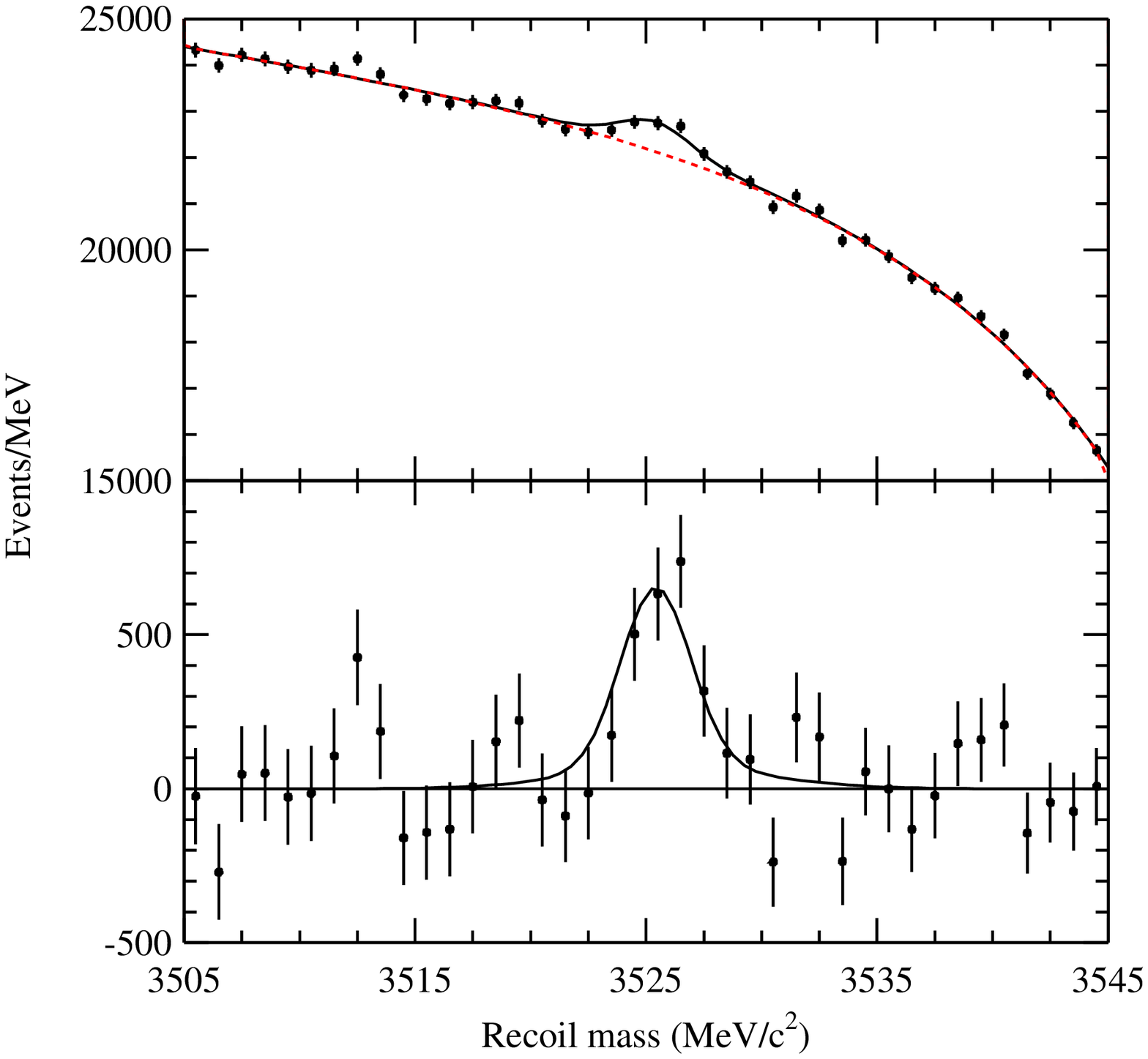}
\end{center}
\caption{Recoil mass spectrum $M(X)$ in $\psi(2S) \to \pi^0 X$ with the
restriction $|\cos \alpha| \le 0.5$.  Top:  unsubtracted; bottom:
background-subtracted.
\label{fig:hc}}
\end{figure}

We report here on a new CLEO result \cite{Ge:2011kq}.  The numbers in this
section and the next are preliminary.
In the decay $\psi(2S) \to \pi^0 h_c$ the $\pi^0$ has an energy of 159 MeV.
An important background from the photon in $\psi(2S) \to \gamma \chi_{c2}$
pairing with a random low-energy photon can be suppressed by rejecting very
asymmetric $\pi^0$ decays.  Fig.\ \ref{fig:egam} shows the dependence on
effective recoil mass against a $\pi^0$ candidate of the energy of the
higher-energy photon for various values of $|\cos \alpha|$, where $\alpha$ is
the angle of either photon in the $\pi^0$ rest frame relative to the $\pi^0$
boost direction.  For values of $|\cos \alpha|$ exceeding about 0.5, there is
danger of confusion of the higher-energy photon in $\pi^0$ with a photon
from $\psi(2S) \to \gamma \chi_{c2}$ (the horizontal dash-dotted line and
the horizontal dotted lines within 6 MeV of it).  Thus, $\pi^0$ candidates
were required to have $|\cos \alpha| < 0.5$.  A preliminary mass spectrum
is shown in Fig.\ \ref{fig:hc}.


The fit in Fig.\ \ref{fig:hc} is performed with $M(h_c)$ fixed to its
world average \cite{Nakamura:2010zzi}.  The branching fraction is found to be
${\cal B}[\psi(2S) \to \pi^0 h_c] = (9.0 \pm 1.5 \pm 1.2) \times 10^{-4}$,
based on (25.89$\pm$0.52) million $\psi(2S)$ decays.  This is to be compared
with the BESIII value \cite{Ablikim:2010rc} of $(8.4 \pm 1.3 \pm 1.0) \times
10^{-4}$ with 106 million $\psi(2S)$ decays.

\section{$\Upsilon(3S) \to \pi^0 h_b$ search \label{sec:hb}}

\begin{figure}
\includegraphics[width=0.48\textwidth]{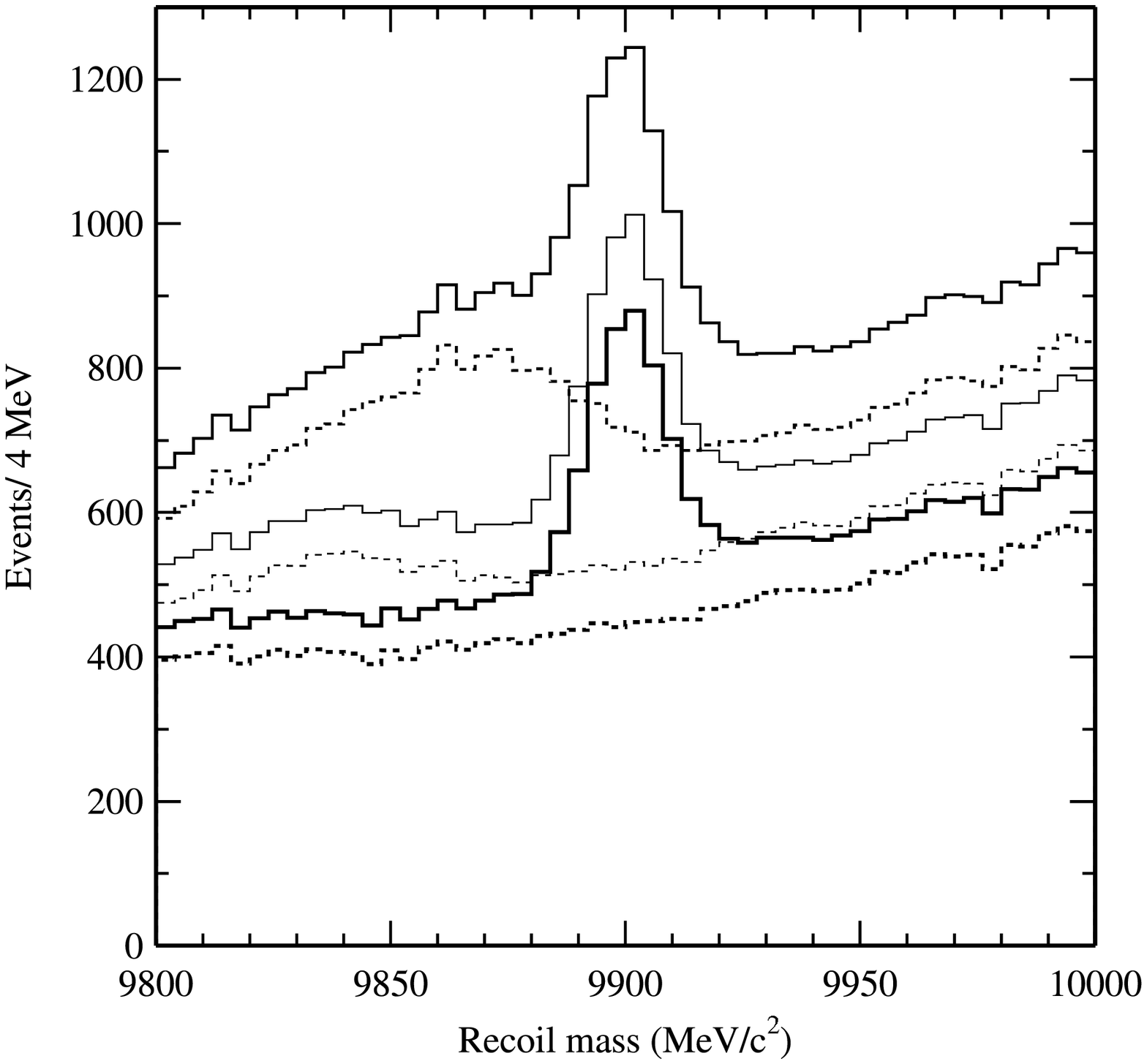}
\includegraphics[width=0.488\textwidth]{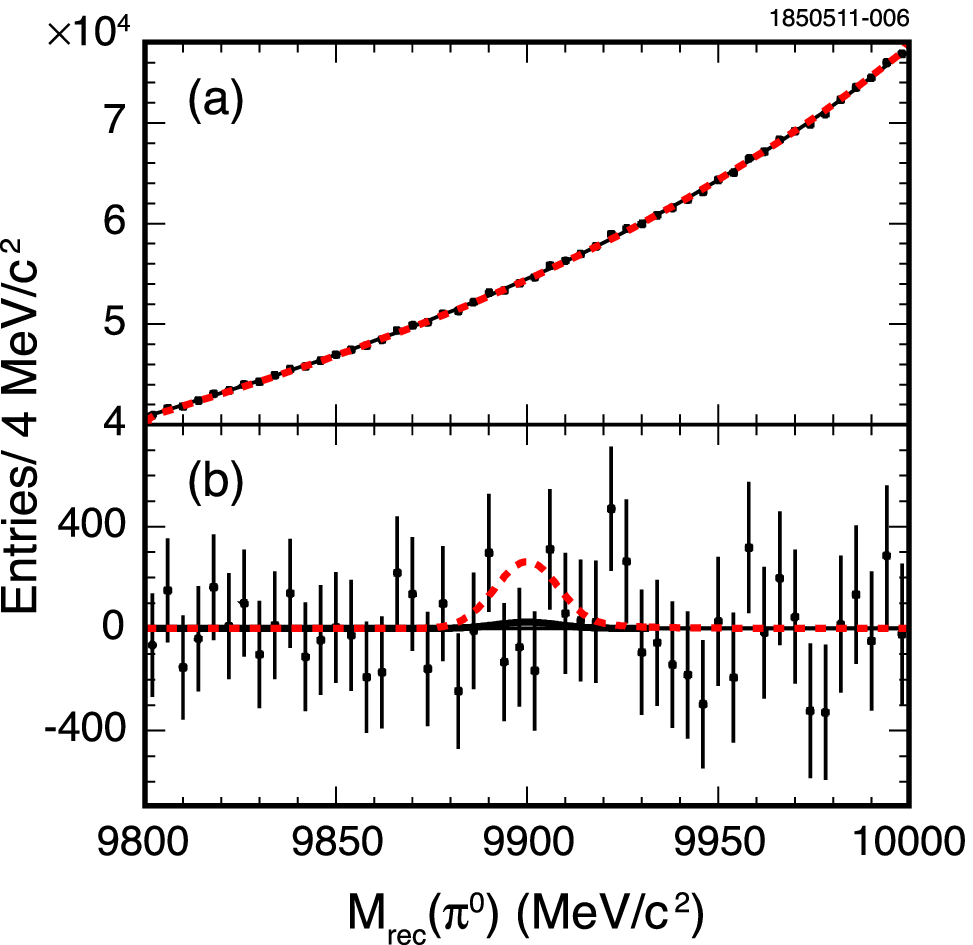}
\caption{Left:  Solid curves show Monte Carlo for $\Upsilon(3S) \to \gamma
\chi_{bJ}(1P)$ and $\Upsilon(3S) \to \pi^0 h_b$ signal Monte Carlo for,
top to bottom, $|\cos \alpha| < 1.0, 0.8$, and 0.7. Dashes denote corresponding
curves for $\chi_{bJ}$ Monte Carlo alone.  Right: (a) Fit to mass recoiling
against a $\pi^0$ for $\Upsilon(3S) \to \pi^0 h_b$ with $M(h_b)$ fixed at 9900
MeV/$c^2$.  (b) Fitted background-subtracted spectrum (solid curve).  The
dashed curve corresponds to the upper limit on signal candidates at 90\% c.l.
\label{fig:hblim}}
\end{figure}

A selection of events with a maximum value of $|\cos \alpha| < 0.7$ was
found to smooth the background due to $\Upsilon(3S) \to \gamma \chi_{bJ}(1P)$
sufficiently (see the left-hand panel of Fig.\ \ref{fig:hblim}) so that the
tightest possible upper limit could be placed on ${\cal B}[\Upsilon(3S) \to
\pi^0 h_b]$.  Choosing the mass range 9895 MeV/$c^2 \le M(h_b) \le 9905$
MeV/$c^2$, an upper limit ${\cal B}[\Upsilon(3S) \to \pi^0 h_b] < 1.2 \times
10^{-3}$ at 90\% c.l.\ was placed \cite{Ge:2011kq} (see the right-hand panel of
Fig.\ \ref{fig:hblim}), superseding a 
previous CLEO 90\% c.l.\ bound \cite{Butler:1993rq} of $2.7 \times 10^{-3}$.
 
\section{Observation of $e^+ e^- \to \pi^+ \pi^- h_c$ at $\sqrt{s} = 4170$
MeV \label{sec:hcpr}}

BaBar \cite{Aubert:2006bm,Aubert:2008bv} and Belle \cite{Sokolov:2006sd} have
observed $\pi^+ \pi^-$ and $\eta$ transitions from $\Upsilon(4S)$ to lower
states; Belle \cite{Abe:2007tk} saw $\pi^+ \pi^-$ transitions from
$\Upsilon(5S)$  to lower $\Upsilon$ states with rates more than 100 times the
$nS$ rates for $n \le 4$ \cite{Rosner:2011}.  This led to the question of
whether any hadronic transitions could be seen to $c \bar c$ states below
flavor threshold from those above it.

\begin{figure}
\includegraphics[width=0.96\textwidth]{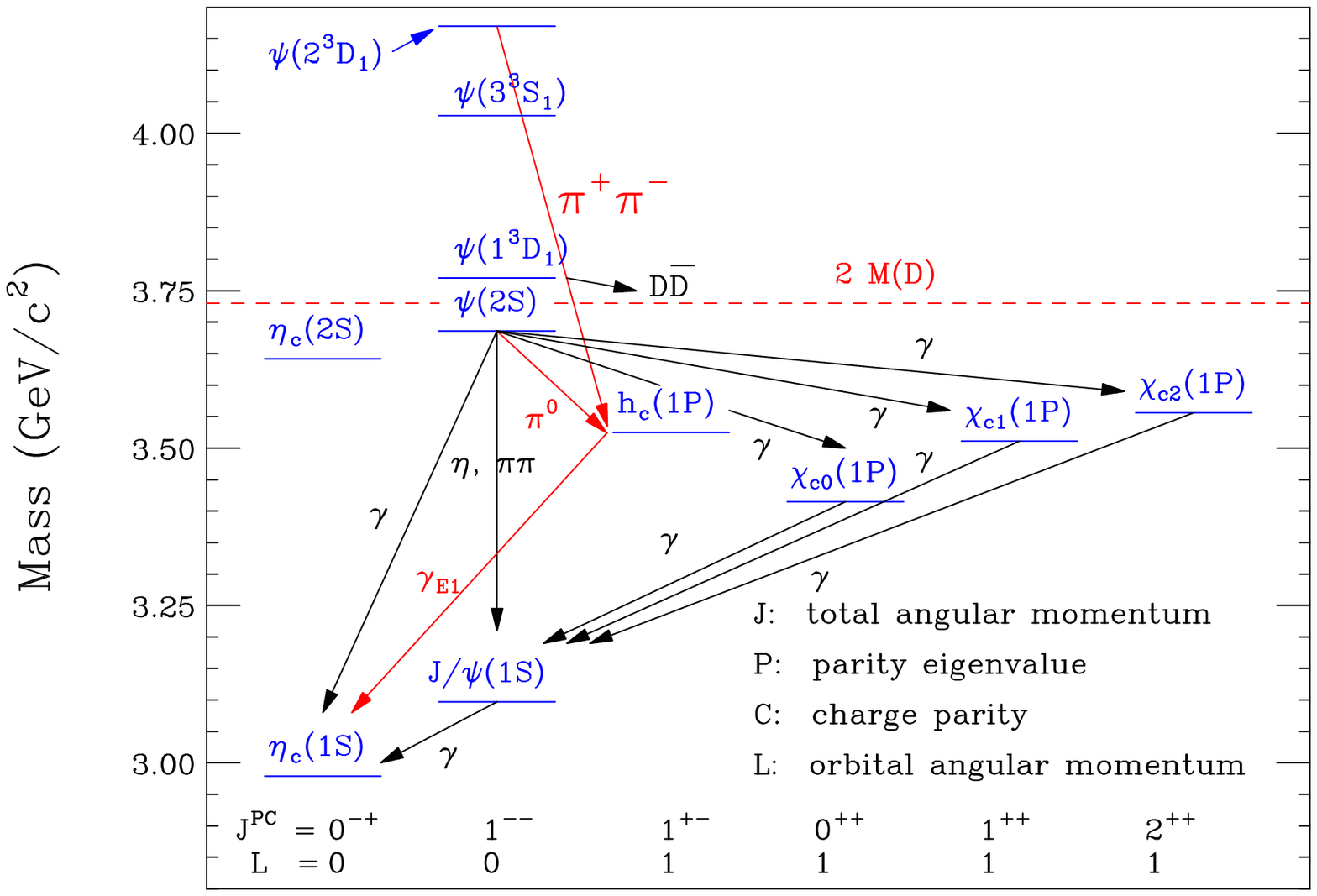}
\caption{Charmonium spectrum.  Red arrows denote transitions of interest:
$\psi(2^3D_1) = \psi(4170) \to \pi^+ \pi^- h_c(1P)$, $\psi(2S) \to \pi^0
h_c(1P)$ (the normalizing transition), and $h_c(1P) \to \gamma \eta_c(1S)$
via an electric dipole (E1) transition.
\label{fig:pphc}}
\end{figure}

\begin{figure}
\includegraphics[width=0.48\textwidth]{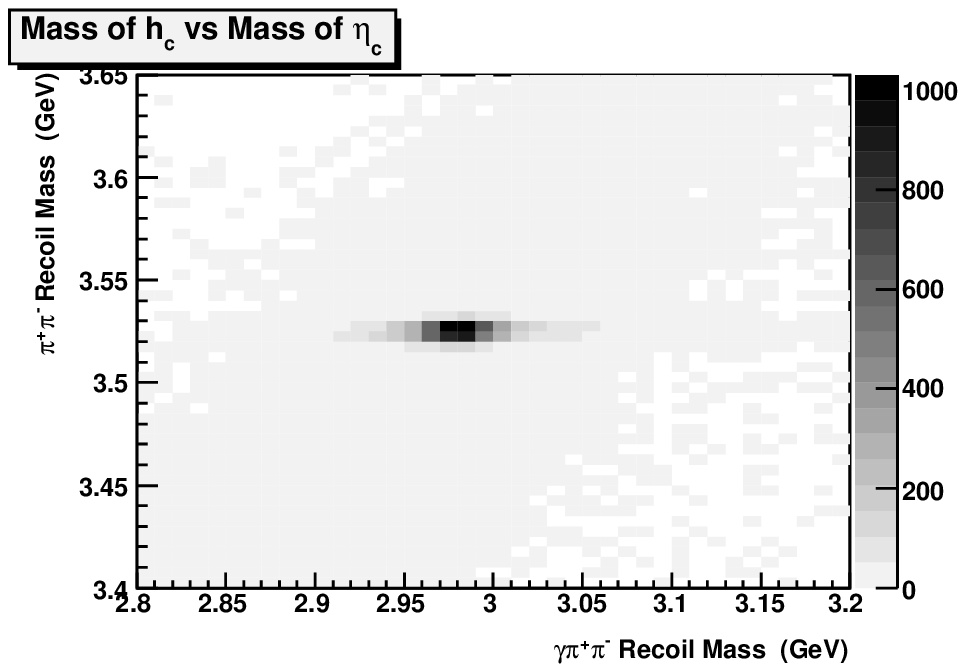}\\
\mbox{\includegraphics[width=0.48\textwidth]{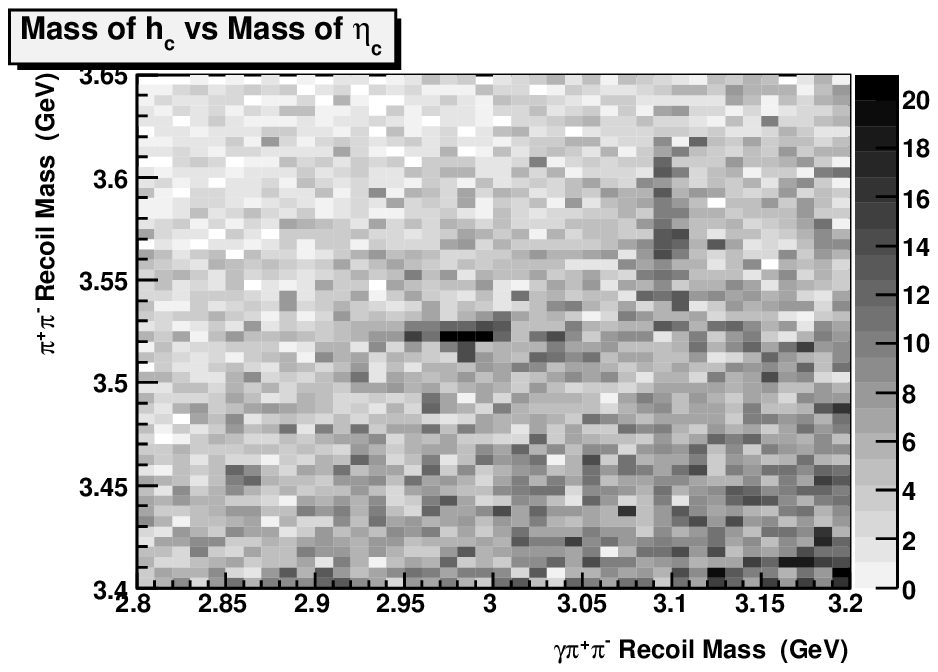}
      \includegraphics[width=0.48\textwidth]{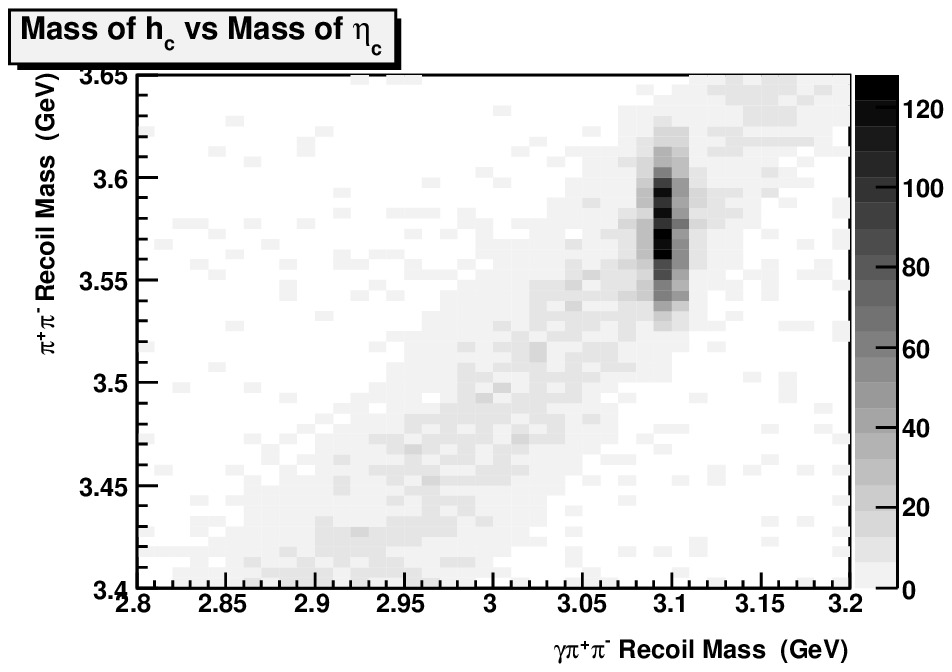}}
\caption{$\pi^+ \pi^-$ recoil mass vs.\ $\gamma \pi^+ \pi^-$ recoil mass.
Upper left:  signal Monte Carlo; lower left:  data (CLEO); lower right:
initial state radiation Monte Carlo.  Note that the data shows evidence for
both the $h_c$ signal and the production of $\psi(2S) \pi^+ \pi^- J/\psi$
through the initial-state radiation process $e^+ e^- \to \gamma \psi(2S)$.
\label{fig:hcsig}}
\end{figure}

Having a large sample of $e^+ e^-$ annihilations at the center-of-mass
energy of 4170 MeV (the approximate energy of the $\psi(2^3D_1)$ candidate),
CLEO searched for and found the transition $\psi(4170) \to \pi^+ \pi^-
h_c(1P)$ \cite{Pedlar:2011uqa}, followed by $h_c(1P) \to \gamma \eta_c(1S)$
with $\eta_c(1S)$ decaying in twelve different exclusive modes.  The transition
rate was normalized by comparing with the known rate \cite{Ablikim:2010rc} for
$\psi(2S) \to \pi^0 h_c(1P)$.  In Fig.\ \ref{fig:pphc} we denote transitions of
interest by red arrows.  Evidence for the signal is shown in the lower left
panel of Fig.\ \ref{fig:hcsig}.

\begin{figure}
\mbox{\includegraphics[width=0.48\textwidth]{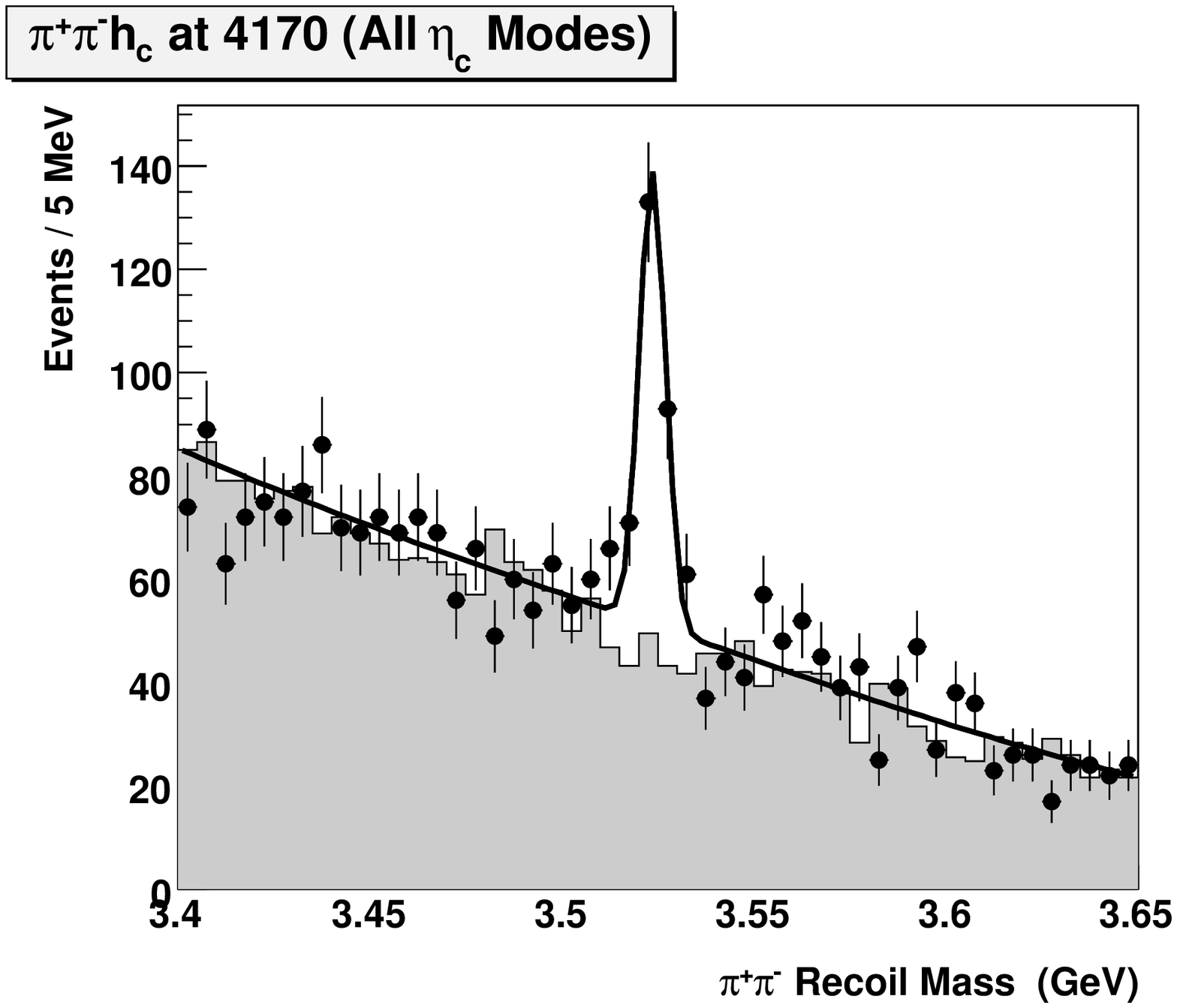}
      \includegraphics[width=0.48\textwidth]{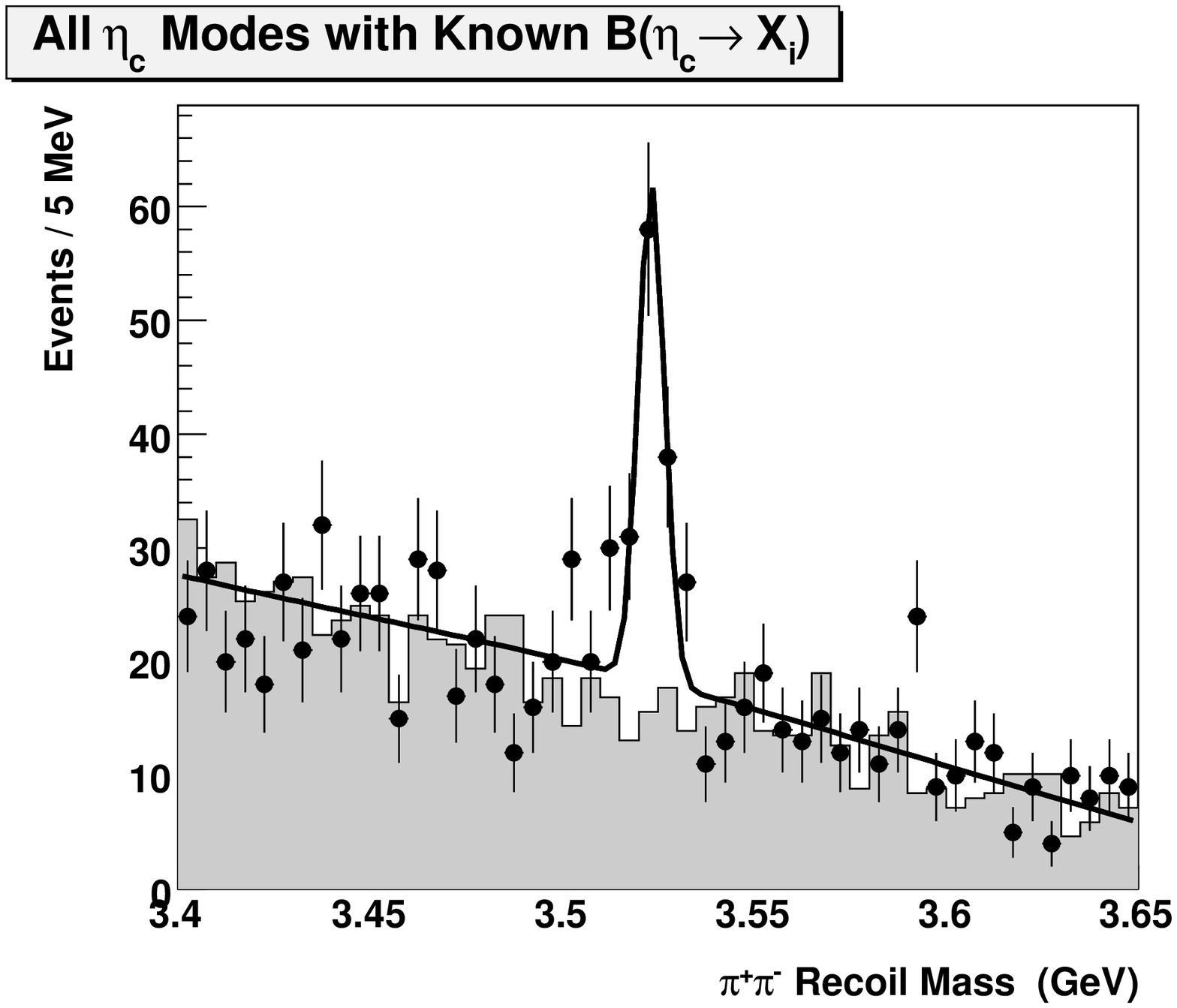}}
\caption{Number of events per 5 MeV vs.\ mass $M(X)$ recoiling against $\pi^+
\pi^-$ in $\psi(4170) \to \pi^+ \pi^- X$.  Left:  all $\eta_c$ modes; 150$\pm$17
events with 9.4$\sigma$ significance; right: modes with known ${\cal B}(\eta_c
\to X_i)$: 74$\pm$11 events with 7.7$\sigma$ significance.
\label{fig:hcpk}}
\end{figure}

One selects the $\eta_c$ using the $\gamma \pi^+ \pi^-$ recoil mass and plots
the $\pi^+ \pi^-$ recoil mass.  Signal is defined by events with kinematic
fit $\chi^2$ per degree of freedom (d.o.f.) less than 5; background is
defined by events with fit $\chi^2/$d.o.f.\ between 10 and 25.  The $h_c$
peak is shown in Fig.\ \ref{fig:hcpk}, on the left for events where $\eta_c$
decays to all twelve chosen modes and on the right for events in which
$\eta_c$ decays to modes for which its branching fraction is known.
The rate for $\psi(4170) \to \pi^+ \pi^- h_c$ is found comparable to that for
$\psi(4170) \to \pi^+ \pi^- J/\psi$, which is curious because the former
process involves a charmed-quark spin flip while the latter does not.
Signals for $e^+ e^- \to (\pi^0\pi^0/\pi^0/\eta)h_c$ at 4170 MeV were searched
for as well but none found significant.

\section{$D^+ \to X \ell \nu$; $D_s^* \to D_s e^+ e^-$; $D \to h e e$
\label{sec:sl}}

CLEO has recently analyzed the decays $D^+ \to \{ \eta',\eta,\phi \} e^+ \nu_e$
\cite{Yelton:2010js}, finding ${\cal B}(D^+ \to \eta' e^+ \nu_e) = (2.16 \pm
0.53 \pm 0.07) \times 10^{-4}$ (a first measurement) and ${\cal B}(D^+ \to \eta
e^+ \nu_e) = (11.4 \pm 0.9 \pm 0.4) \times 10^{-4}$ (a first measurement of the
form factor).  Comparing the two rates sheds light on the nonstrange quark
content of the $\eta$ and $\eta'$.  An upper bound ${\cal B}(D^+ \to \phi e^+
\nu_e) < 0.9 \times 10^{-4}$ (90\% c.l.) was placed in this same work.

The Dalitz decay $D_s^{*+} \to D_s^+ e^+ e^-$ has been observed by CLEO for
the first time \cite{CroninHennessy:2011xp}.  The result is ${\cal B}(D_s^{*+}
\to D_s e^+ e^-)/ {\cal B}(D_s^{*+} \to D_s \gamma) = (0.72^{+0.15}_{-0.13}
\pm 0.10)\%$, in accord with Standard Model predictions.

Limits have been placed by CLEO on a variety of $D_{(s)} \to h^\pm e e$
processes \cite{Rubin:2010cq}.  These are summarized in Table \ref{tab:hee}.


\begin{table}
\caption{Summary of CLEO limits \cite{Rubin:2010cq} on $D_{(s)} \to h^\pm e e$
processes.
\label{tab:hee}}
\begin{center}
\begin{tabular}{c c c c c} \hline \hline
Decay of: & \multicolumn{2}{c}{$D^+$} & \multicolumn{2}{c}{$D_s^+$} \\
$ee=$~~~$h = $ & $\pi$ & $K$ & $\pi$ & $K$ \\ \hline
$e^+ e^-$~~~~~ & $5.9 \times 10^{-6}$ & $3.0 \times 10^{-6}$
          & $2.2 \times 10^{-5}$ & $5.2 \times 10^{-5}$ \\
$e^+ e^+$~~~~~ & $1.1 \times 10^{-6}$ & $3.5 \times 10^{-6}$
          & $1.8 \times 10^{-5}$ & $1.7 \times 10^{-5}$ \\ \hline \hline
\end{tabular}
\end{center}
\end{table}

\section{$D_s^+ \to \omega e^+ \nu_e$ bound \label{sec:omenu}}

\begin{figure}
\includegraphics[width=0.55\textwidth]{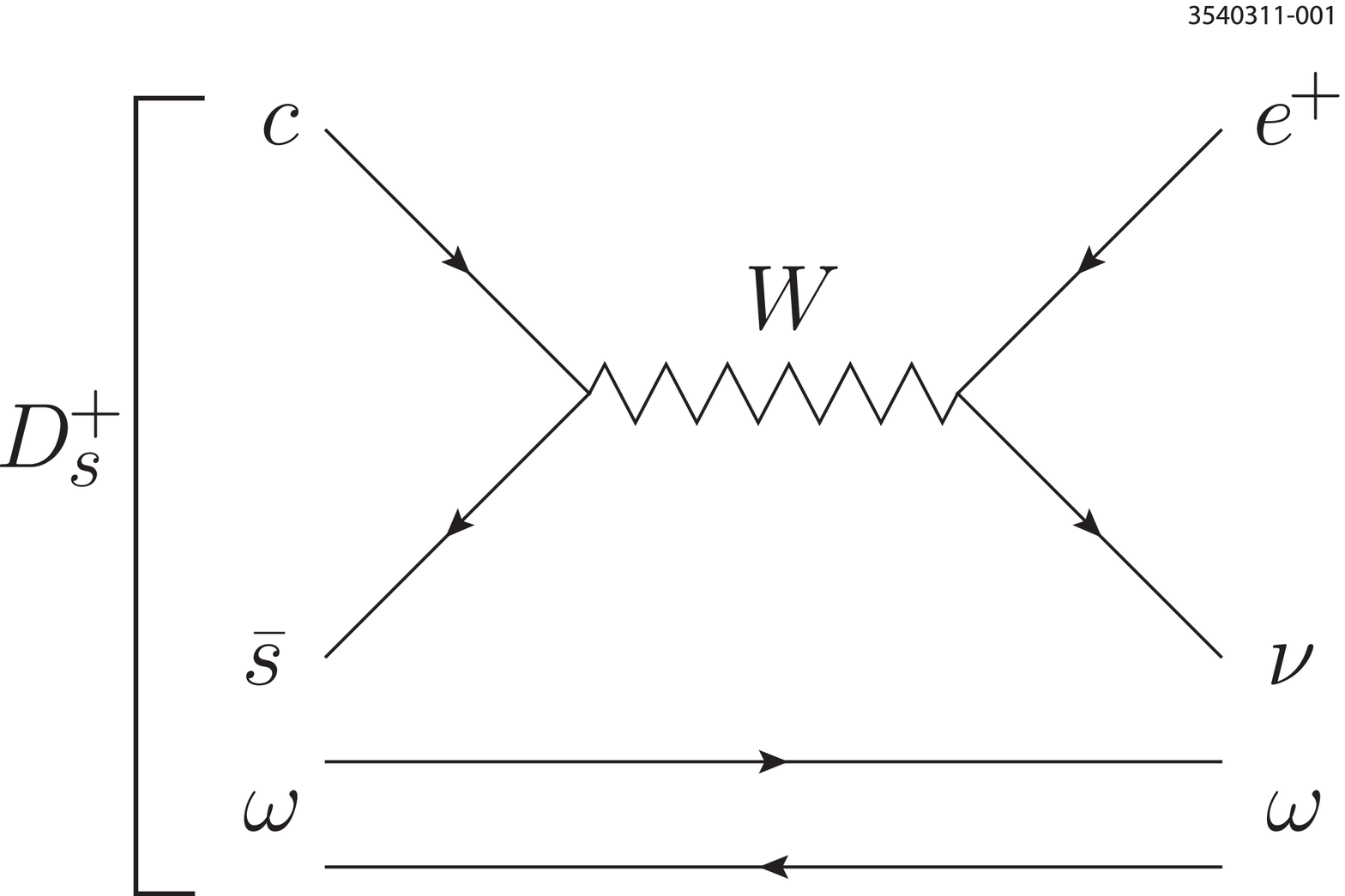}
\includegraphics[width=0.39\textwidth]{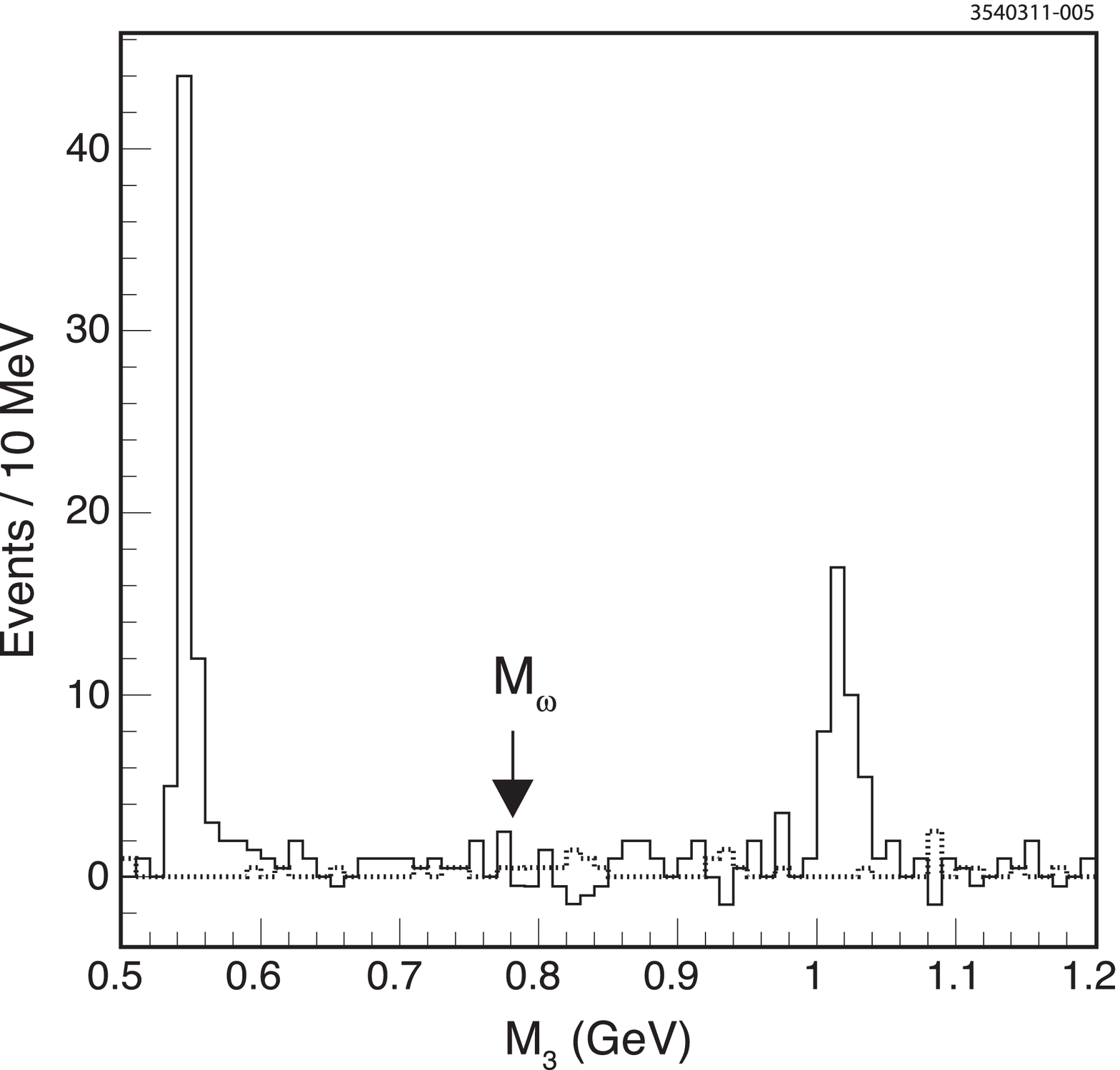}
\caption{Left:  Example of a Feynman diagram contributing to $D_s^+ \to \omega
e^+ \nu_e$ \cite{Martin:2011rd}.  Right:  $(\pi^+ \pi^- \pi^0)$ spectrum $M_3$
in $D_s^+ \to \pi^+ \pi^- \pi^0 e^+ \nu_e$ \cite{Martin:2011rd}.
\label{fig:omenu}}
\end{figure}

$D_s$ Cabibbo-favored semileptonic decays are expected to lead to final
states which can couple to $s \bar s$; the $\omega \simeq (u \bar u + d \bar d)
/\sqrt{2}$ has none.  Hence the decay $D_s^+ \to \omega e^+ \nu_3$ tests
mixing or a phenomenon known as ``weak annihilation'' (WA) \cite{Gronau:2009mp}.
One diagram contributing to such a process is shown in the left-hand panel
of Fig.\ \ref{fig:omenu}.
A search for this process has been performed by CLEO \cite{Martin:2011rd}, with
the resulting branching fraction ${\cal B}(D_s^+ \to \omega e^+ \nu_e) < 0.20
\%$ at 90\% c.l.  As shown in the right-hand panel of Fig.\ \ref{fig:omenu},
one sees evidence for $D_s^+ \to \eta e^+ \nu_e$ and $D_s^+ \to \phi e^+ \nu_e$
in the three-pion $(\pi^+ \pi^- \pi^0)$ spectrum $M_3$, but not $D_s^+ \to
\omega e^+ \nu_e$.  In Ref.\ \cite{Gronau:2009mp} this branching fraction was
estimated to be $(0.13 \pm 0.05)\%$ by assuming $D_s^+ \to \omega \pi^+$ is
dominated by WA and using factorization, and
to be no more than $2 \times 10^{-4}$ if due to mixing alone.

\section{Conclusions \label{sec:concl}}

CLEO is continuing to contribute to charm (and bottom) physics although it
ceased running more than three years ago.  Its large sample of correlated $D$
and $\bar D$ mesons produced at $\psi(3770)$ can produce information about
relative strong phases in Dalitz plots for $D$ decays, useful in studies of
CP-violating $B$ decays.  It has improved knowledge of radiative transitions
involving the lowest P-wave $b \bar b$ states and measured some suppressed
branching fractions for the first time.   The transition $\psi(4170) \to \pi^+
\pi^- h_c$, with $h_c \to \gamma \eta_c$, represents an early observation of
a hadronic transition from a quarkonium state above flavor threshold to one
below it, possibly indicating the importance of rescattering from flavored
meson-antimeson pairs.  Numerous $D$ and $D_s$ decays (semileptonic,
Dalitz, OZI-suppressed) are still being mapped out.

Many questions may require methods beyond the capability of current theoretical
approaches.  For example, it is easy to pose a problem (``too many quarks'')
that lattice QCD can't handle.  We are still learning about ($c$,$b$) mesons
after more than thirty years.  Many potential tests of our understanding of
low-energy strong interactions remain to be performed.

\Acknowledgements
This work was supported in part by the United States Department of Energy
under Grant No.\ DE-FG02-90ER40560.

\end{document}